\documentclass[12pt,preprint]{aastex}
\newcommand{\etal}{\hbox{ et~al.}}
\newcommand{\eg}{\hbox{e.g.}}

\def\PsfigVersion{1.10}
\def\setDriver{\DvipsDriver} 
\ifx\undefined\psfig\else \fi
\let\LaTeXAtSign=\@
\let\@=\relax
\edef\psfigRestoreAt{\catcode`\@=\number\catcode`@\relax}
\catcode`\@=11\relax
\newwrite\@unused
\def\ps@typeout#1{{\let\protect\string\immediate\write\@unused{#1}}}

\def\DvipsDriver{
	\ps@typeout{psfig/tex \PsfigVersion -dvips}
\def\PsfigSpecials{\DvipsSpecials} 	\def\ps@dir{/}
\def\ps@predir{} }
\def\OzTeXDriver{
	\ps@typeout{psfig/tex \PsfigVersion -oztex}
	\def\PsfigSpecials{\OzTeXSpecials}
	\def\ps@dir{:}
	\def\ps@predir{:}
	\catcode`\^^J=5
}

\def\figurepath{./:}

\def\DoPaths#1{\expandafter\EachPath#1\stoplist}
\def\leer{}
\def\EachPath#1:#2\stoplist{
  \ExistsFile{#1}{\SearchedFile}
  \ifx#2\leer
  \else
    \expandafter\EachPath#2\stoplist
  \fi}
\def\ps@dir{/}
\def\ExistsFile#1#2{%
   \openin1=\ps@predir#1\ps@dir#2
   \ifeof1
       \closein1
   \else
       \closein1
        \ifx\ps@founddir\leer
           \edef\ps@founddir{#1}
        \fi
   \fi}
\def\get@dir#1{%
  \def\ps@founddir{}
  \def\SearchedFile{#1}
  \DoPaths\figurepath
}

\def\@nnil{\@nil}
\def\@empty{}
\def\@psdonoop#1\@@#2#3{}
\def\@psdo#1:=#2\do#3{\edef\@psdotmp{#2}\ifx\@psdotmp\@empty \else
    \expandafter\@psdoloop#2,\@nil,\@nil\@@#1{#3}\fi}
\def\@psdoloop#1,#2,#3\@@#4#5{\def#4{#1}\ifx #4\@nnil \else
       #5\def#4{#2}\ifx #4\@nnil \else#5\@ipsdoloop #3\@@#4{#5}\fi\fi}
\def\@ipsdoloop#1,#2\@@#3#4{\def#3{#1}\ifx #3\@nnil 
       \let\@nextwhile=\@psdonoop \else
      #4\relax\let\@nextwhile=\@ipsdoloop\fi\@nextwhile#2\@@#3{#4}}
\def\@tpsdo#1:=#2\do#3{\xdef\@psdotmp{#2}\ifx\@psdotmp\@empty \else
    \@tpsdoloop#2\@nil\@nil\@@#1{#3}\fi}
\def\@tpsdoloop#1#2\@@#3#4{\def#3{#1}\ifx #3\@nnil 
       \let\@nextwhile=\@psdonoop \else
      #4\relax\let\@nextwhile=\@tpsdoloop\fi\@nextwhile#2\@@#3{#4}}
\ifx\undefined\fbox
\newdimen\fboxrule
\newdimen\fboxsep
\newdimen\ps@tempdima
\newbox\ps@tempboxa
\long\def\fbox#1{\leavevmode\setbox\ps@tempboxa\hbox{#1}\ps@tempdima\fboxrule
    \advance\ps@tempdima \fboxsep \advance\ps@tempdima \dp\ps@tempboxa
   \hbox{\lower \ps@tempdima\hbox
  {\vbox{\hrule height \fboxrule
          \hbox{\vrule width \fboxrule \hskip\fboxsep
          \vbox{\vskip\fboxsep \box\ps@tempboxa\vskip\fboxsep}\hskip 
                 \fboxsep\vrule width \fboxrule}
                 \hrule height \fboxrule}}}}
\fi
\newread\ps@stream
\newif\ifnot@eof       
\newif\if@noisy        
\newif\if@atend        
\newif\if@psfile       
{\catcode`\%=12\global\gdef\epsf@start{
\def\epsf@PS{PS}
\def\epsf@getbb#1{%
\openin\ps@stream=\ps@predir#1
\ifeof\ps@stream\ps@typeout{Error, File #1 not found}\else
   {\not@eoftrue \chardef\other=12
    \def\do##1{\catcode`##1=\other}\dospecials \catcode`\ =10
    \loop
       \if@psfile
	  \read\ps@stream to \epsf@fileline
       \else{
	  \obeyspaces
          \read\ps@stream to \epsf@tmp\global\let\epsf@fileline\epsf@tmp}
       \fi
       \ifeof\ps@stream\not@eoffalse\else
       \if@psfile\else
       \expandafter\epsf@test\epsf@fileline:. \\%
       \fi
          \expandafter\epsf@aux\epsf@fileline:. \\%
       \fi
   \ifnot@eof\repeat
   }\closein\ps@stream\fi}%
\long\def\epsf@test#1#2#3:#4\\{\def\epsf@testit{#1#2}
			\ifx\epsf@testit\epsf@start\else
\ps@typeout{Warning! File does not start with `\epsf@start'.  It may not be a PostScript file.}
			\fi
			\@psfiletrue} 
{\catcode`\%=12\global\let\epsf@percent=
\long\def\epsf@aux#1#2:#3\\{\ifx#1\epsf@percent
   \def\epsf@testit{#2}\ifx\epsf@testit\epsf@bblit
	\@atendfalse
        \epsf@atend #3 . \\%
	\if@atend	
	   \if@verbose{
		\ps@typeout{psfig: found `(atend)'; continuing search}
	   }\fi
        \else
        \epsf@grab #3 . . . \\%
        \not@eoffalse
        \global\no@bbfalse
        \fi
   \fi\fi}%
\def\epsf@grab #1 #2 #3 #4 #5\\{%
   \global\def\epsf@llx{#1}\ifx\epsf@llx\empty
      \epsf@grab #2 #3 #4 #5 .\\\else
   \global\def\epsf@lly{#2}%
   \global\def\epsf@urx{#3}\global\def\epsf@ury{#4}\fi}%
\def\epsf@atendlit{(atend)} 
\def\epsf@atend #1 #2 #3\\{%
   \def\epsf@tmp{#1}\ifx\epsf@tmp\empty
      \epsf@atend #2 #3 .\\\else
   \ifx\epsf@tmp\epsf@atendlit\@atendtrue\fi\fi}

\chardef\psletter = 11 
\chardef\other = 12

\newif \ifdebug 
\newif\ifc@mpute 
\c@mputetrue 

\let\then = \relax
\def\r@dian{pt }
\let\r@dians = \r@dian
\let\dimensionless@nit = \r@dian
\let\dimensionless@nits = \dimensionless@nit
\def\internal@nit{sp }
\let\internal@nits = \internal@nit
\newif\ifstillc@nverging
\def \Mess@ge #1{\ifdebug \then \message {#1} \fi}

{ 
	\catcode `\@ = \psletter
	\gdef \nodimen {\expandafter \n@dimen \the \dimen}
	\gdef \term #1 #2 #3%
	       {\edef \t@ {\the #1}
		\edef \t@@ {\expandafter \n@dimen \the #2\r@dian}%
		\t@rm {\t@} {\t@@} {#3}%
	       }
	\gdef \t@rm #1 #2 #3%
	       {{%
		\count 0 = 0
		\dimen 0 = 1 \dimensionless@nit
		\dimen 2 = #2\relax
		\Mess@ge {Calculating term #1 of \nodimen 2}%
		\loop
		\ifnum	\count 0 < #1
		\then	\advance \count 0 by 1
			\Mess@ge {Iteration \the \count 0 \space}%
			\Multiply \dimen 0 by {\dimen 2}%
			\Mess@ge {After multiplication, term = \nodimen 0}%
			\Divide \dimen 0 by {\count 0}%
			\Mess@ge {After division, term = \nodimen 0}%
		\repeat
		\Mess@ge {Final value for term #1 of 
				\nodimen 2 \space is \nodimen 0}%
		\xdef \Term {#3 = \nodimen 0 \r@dians}%
		\aftergroup \Term
	       }}
	\catcode `\p = \other
	\catcode `\t = \other
	\gdef \n@dimen #1pt{#1} 
}

\def \Divide #1by #2{\divide #1 by #2} 

\def \Multiply #1by #2
       {{
	\count 0 = #1\relax
	\count 2 = #2\relax
	\count 4 = 65536
	\Mess@ge {Before scaling, count 0 = \the \count 0 \space and
			count 2 = \the \count 2}%
	\ifnum	\count 0 > 32767 
	\then	\divide \count 0 by 4
		\divide \count 4 by 4
	\else	\ifnum	\count 0 < -32767
		\then	\divide \count 0 by 4
			\divide \count 4 by 4
		\else
		\fi
	\fi
	\ifnum	\count 2 > 32767 
	\then	\divide \count 2 by 4
		\divide \count 4 by 4
	\else	\ifnum	\count 2 < -32767
		\then	\divide \count 2 by 4
			\divide \count 4 by 4
		\else
		\fi
	\fi
	\multiply \count 0 by \count 2
	\divide \count 0 by \count 4
	\xdef \product {#1 = \the \count 0 \internal@nits}%
	\aftergroup \product
       }}

\def\r@duce{\ifdim\dimen0 > 90\r@dian \then   
		\multiply\dimen0 by -1
		\advance\dimen0 by 180\r@dian
		\r@duce
	    \else \ifdim\dimen0 < -90\r@dian \then  
		\advance\dimen0 by 360\r@dian
		\r@duce
		\fi
	    \fi}

\def\Sine#1%
       {{%
	\dimen 0 = #1 \r@dian
	\r@duce
	\ifdim\dimen0 = -90\r@dian \then
	   \dimen4 = -1\r@dian
	   \c@mputefalse
	\fi
	\ifdim\dimen0 = 90\r@dian \then
	   \dimen4 = 1\r@dian
	   \c@mputefalse
	\fi
	\ifdim\dimen0 = 0\r@dian \then
	   \dimen4 = 0\r@dian
	   \c@mputefalse
	\fi
	\ifc@mpute \then
		\divide\dimen0 by 180
		\dimen0=3.141592654\dimen0
		\dimen 2 = 3.1415926535897963\r@dian 
		\divide\dimen 2 by 2 
		\Mess@ge {Sin: calculating Sin of \nodimen 0}%
		\count 0 = 1 
		\dimen 2 = 1 \r@dian 
		\dimen 4 = 0 \r@dian 
		\loop
			\ifnum	\dimen 2 = 0 
			\then	\stillc@nvergingfalse 
			\else	\stillc@nvergingtrue
			\fi
			\ifstillc@nverging 
			\then	\term {\count 0} {\dimen 0} {\dimen 2}%
				\advance \count 0 by 2
				\count 2 = \count 0
				\divide \count 2 by 2
				\ifodd	\count 2 
				\then	\advance \dimen 4 by \dimen 2
				\else	\advance \dimen 4 by -\dimen 2
				\fi
		\repeat
	\fi		
			\xdef \sine {\nodimen 4}%
       }}

\def\Cosine#1{\ifx\sine\UnDefined\edef\Savesine{\relax}\else
		             \edef\Savesine{\sine}\fi
	{\dimen0=#1\r@dian\advance\dimen0 by 90\r@dian
	 \Sine{\nodimen 0}
	 \xdef\cosine{\sine}
	 \xdef\sine{\Savesine}}}	      

\def\psdraft{
	\def\@psdraft{0}
}
\def\psfull{
	\def\@psdraft{100}
}

\psfull

\newif\if@scalefirst
\def\psscalefirst{\@scalefirsttrue}
\def\psrotatefirst{\@scalefirstfalse}
\psrotatefirst

\newif\if@draftbox
\def\psnodraftbox{
	\@draftboxfalse
}
\def\psdraftbox{
	\@draftboxtrue
}
\@draftboxtrue

\newif\if@prologfile
\newif\if@postlogfile
\def\pssilent{
	\@noisyfalse
}
\def\psnoisy{
	\@noisytrue
}
\psnoisy
\newif\if@bbllx
\newif\if@bblly
\newif\if@bburx
\newif\if@bbury
\newif\if@height
\newif\if@width
\newif\if@rheight
\newif\if@rwidth
\newif\if@angle
\newif\if@clip
\newif\if@verbose
\def\@p@@sclip#1{\@cliptrue}
\newif\if@decmpr
\def\@p@@sfigure#1{\def\@p@sfile{null}\def\@p@sbbfile{null}\@decmprfalse
   \openin1=\ps@predir#1
   \ifeof1
	\closein1
	\get@dir{#1}
	\ifx\ps@founddir\leer
		\openin1=\ps@predir#1.bb
		\ifeof1
			\closein1
			\get@dir{#1.bb}
			\ifx\ps@founddir\leer
				\ps@typeout{Can't find #1 in \figurepath}
			\else
				\@decmprtrue
				\def\@p@sfile{\ps@founddir\ps@dir#1}
				\def\@p@sbbfile{\ps@founddir\ps@dir#1.bb}
			\fi
		\else
			\closein1
			\@decmprtrue
			\def\@p@sfile{#1}
			\def\@p@sbbfile{#1.bb}
		\fi
	\else
		\def\@p@sfile{\ps@founddir\ps@dir#1}
		\def\@p@sbbfile{\ps@founddir\ps@dir#1}
	\fi
   \else
	\closein1
	\def\@p@sfile{#1}
	\def\@p@sbbfile{#1}
   \fi
}
\def\@p@@sfile#1{\@p@@sfigure{#1}}
\def\@p@@sbbllx#1{
		\@bbllxtrue
		\dimen100=#1
		\edef\@p@sbbllx{\number\dimen100}
}
\def\@p@@sbblly#1{
		\@bbllytrue
		\dimen100=#1
		\edef\@p@sbblly{\number\dimen100}
}
\def\@p@@sbburx#1{
		\@bburxtrue
		\dimen100=#1
		\edef\@p@sbburx{\number\dimen100}
}
\def\@p@@sbbury#1{
		\@bburytrue
		\dimen100=#1
		\edef\@p@sbbury{\number\dimen100}
}
\def\@p@@sheight#1{
		\@heighttrue
		\dimen100=#1
   		\edef\@p@sheight{\number\dimen100}
}
\def\@p@@swidth#1{
		\@widthtrue
		\dimen100=#1
		\edef\@p@swidth{\number\dimen100}
}
\def\@p@@srheight#1{
		\@rheighttrue
		\dimen100=#1
		\edef\@p@srheight{\number\dimen100}
}
\def\@p@@srwidth#1{
		\@rwidthtrue
		\dimen100=#1
		\edef\@p@srwidth{\number\dimen100}
}
\def\@p@@sangle#1{
		\@angletrue
		\edef\@p@sangle{#1} 
}
\def\@p@@ssilent#1{ 
		\@verbosefalse
}
\def\@p@@sprolog#1{\@prologfiletrue\def\@prologfileval{#1}}
\def\@p@@spostlog#1{\@postlogfiletrue\def\@postlogfileval{#1}}
\def\@cs@name#1{\csname #1\endcsname}
\def\@setparms#1=#2,{\@cs@name{@p@@s#1}{#2}}
\def\ps@init@parms{
		\@bbllxfalse \@bbllyfalse
		\@bburxfalse \@bburyfalse
		\@heightfalse \@widthfalse
		\@rheightfalse \@rwidthfalse
		\def\@p@sbbllx{}\def\@p@sbblly{}
		\def\@p@sbburx{}\def\@p@sbbury{}
		\def\@p@sheight{}\def\@p@swidth{}
		\def\@p@srheight{}\def\@p@srwidth{}
		\def\@p@sangle{0}
		\def\@p@sfile{} \def\@p@sbbfile{}
		\def\@p@scost{10}
		\def\@sc{}
		\@prologfilefalse
		\@postlogfilefalse
		\@clipfalse
		\if@noisy
			\@verbosetrue
		\else
			\@verbosefalse
		\fi
}
\def\parse@ps@parms#1{
	 	\@psdo\@psfiga:=#1\do
		   {\expandafter\@setparms\@psfiga,}}
\newif\ifno@bb
\def\bb@missing{
	\if@verbose{
		\ps@typeout{psfig: searching \@p@sbbfile \space  for bounding box}
	}\fi
	\no@bbtrue
	\epsf@getbb{\@p@sbbfile}
        \ifno@bb \else \bb@cull\epsf@llx\epsf@lly\epsf@urx\epsf@ury\fi
}	
\def\bb@cull#1#2#3#4{
	\dimen100=#1 bp\edef\@p@sbbllx{\number\dimen100}
	\dimen100=#2 bp\edef\@p@sbblly{\number\dimen100}
	\dimen100=#3 bp\edef\@p@sbburx{\number\dimen100}
	\dimen100=#4 bp\edef\@p@sbbury{\number\dimen100}
	\no@bbfalse
}
\newdimen\p@intvaluex
\newdimen\p@intvaluey
\def\rotate@#1#2{{\dimen0=#1 sp\dimen1=#2 sp
		  \global\p@intvaluex=\cosine\dimen0
		  \dimen3=\sine\dimen1
		  \global\advance\p@intvaluex by -\dimen3
		  \global\p@intvaluey=\sine\dimen0
		  \dimen3=\cosine\dimen1
		  \global\advance\p@intvaluey by \dimen3
		  }}
\def\compute@bb{
		\no@bbfalse
		\if@bbllx \else \no@bbtrue \fi
		\if@bblly \else \no@bbtrue \fi
		\if@bburx \else \no@bbtrue \fi
		\if@bbury \else \no@bbtrue \fi
		\ifno@bb \bb@missing \fi
		\ifno@bb \ps@typeout{FATAL ERROR: no bb supplied or found}
			\no-bb-error
		\fi
		\count203=\@p@sbburx
		\count204=\@p@sbbury
		\advance\count203 by -\@p@sbbllx
		\advance\count204 by -\@p@sbblly
		\edef\ps@bbw{\number\count203}
		\edef\ps@bbh{\number\count204}
		\if@angle 
			\Sine{\@p@sangle}\Cosine{\@p@sangle}
	        	{\dimen100=\maxdimen\xdef\r@p@sbbllx{\number\dimen100}
					    \xdef\r@p@sbblly{\number\dimen100}
			                    \xdef\r@p@sbburx{-\number\dimen100}
					    \xdef\r@p@sbbury{-\number\dimen100}}
                        \def\minmaxtest{
			   \ifnum\number\p@intvaluex<\r@p@sbbllx
			      \xdef\r@p@sbbllx{\number\p@intvaluex}\fi
			   \ifnum\number\p@intvaluex>\r@p@sbburx
			      \xdef\r@p@sbburx{\number\p@intvaluex}\fi
			   \ifnum\number\p@intvaluey<\r@p@sbblly
			      \xdef\r@p@sbblly{\number\p@intvaluey}\fi
			   \ifnum\number\p@intvaluey>\r@p@sbbury
			      \xdef\r@p@sbbury{\number\p@intvaluey}\fi
			   }
			\rotate@{\@p@sbbllx}{\@p@sbblly}
			\minmaxtest
			\rotate@{\@p@sbbllx}{\@p@sbbury}
			\minmaxtest
			\rotate@{\@p@sbburx}{\@p@sbblly}
			\minmaxtest
			\rotate@{\@p@sbburx}{\@p@sbbury}
			\minmaxtest
			\edef\@p@sbbllx{\r@p@sbbllx}\edef\@p@sbblly{\r@p@sbblly}
			\edef\@p@sbburx{\r@p@sbburx}\edef\@p@sbbury{\r@p@sbbury}
		\fi
		\count203=\@p@sbburx
		\count204=\@p@sbbury
		\advance\count203 by -\@p@sbbllx
		\advance\count204 by -\@p@sbblly
		\edef\@bbw{\number\count203}
		\edef\@bbh{\number\count204}
}
\def\in@hundreds#1#2#3{\count240=#2 \count241=#3
		     \count100=\count240	
		     \divide\count100 by \count241
		     \count101=\count100
		     \multiply\count101 by \count241
		     \advance\count240 by -\count101
		     \multiply\count240 by 10
		     \count101=\count240	
		     \divide\count101 by \count241
		     \count102=\count101
		     \multiply\count102 by \count241
		     \advance\count240 by -\count102
		     \multiply\count240 by 10
		     \count102=\count240	
		     \divide\count102 by \count241
		     \count200=#1\count205=0
		     \count201=\count200
			\multiply\count201 by \count100
		 	\advance\count205 by \count201
		     \count201=\count200
			\divide\count201 by 10
			\multiply\count201 by \count101
			\advance\count205 by \count201
		     \count201=\count200
			\divide\count201 by 100
			\multiply\count201 by \count102
			\advance\count205 by \count201
		     \edef\@result{\number\count205}
}
\def\compute@wfromh{
		\in@hundreds{\@p@sheight}{\@bbw}{\@bbh}
		\edef\@p@swidth{\@result}
}
\def\compute@hfromw{
	        \in@hundreds{\@p@swidth}{\@bbh}{\@bbw}
		\edef\@p@sheight{\@result}
}
\def\compute@handw{
		\if@height 
			\if@width
			\else
				\compute@wfromh
			\fi
		\else 
			\if@width
				\compute@hfromw
			\else
				\edef\@p@sheight{\@bbh}
				\edef\@p@swidth{\@bbw}
			\fi
		\fi
}
\def\compute@resv{
		\if@rheight \else \edef\@p@srheight{\@p@sheight} \fi
		\if@rwidth \else \edef\@p@srwidth{\@p@swidth} \fi
}
\def\compute@sizes{
	\compute@bb
	\if@scalefirst\if@angle
	\if@width
	   \in@hundreds{\@p@swidth}{\@bbw}{\ps@bbw}
	   \edef\@p@swidth{\@result}
	\fi
	\if@height
	   \in@hundreds{\@p@sheight}{\@bbh}{\ps@bbh}
	   \edef\@p@sheight{\@result}
	\fi
	\fi\fi
	\compute@handw
	\compute@resv}
\def\OzTeXSpecials{
	\special{empty.ps /@isp {true} def}
	\special{empty.ps \@p@swidth \space \@p@sheight \space
			\@p@sbbllx \space \@p@sbblly \space
			\@p@sbburx \space \@p@sbbury \space
			startTexFig \space }
	\if@clip{
		\if@verbose{
			\ps@typeout{(clip)}
		}\fi
		\special{empty.ps doclip \space }
	}\fi
	\if@angle{
		\if@verbose{
			\ps@typeout{(rotate)}
		}\fi
		\special {empty.ps \@p@sangle \space rotate \space} 
	}\fi
	\if@prologfile
	    \special{\@prologfileval \space } \fi
	\if@decmpr{
		\if@verbose{
			\ps@typeout{psfig: Compression not available
			in OzTeX version \space }
		}\fi
	}\else{
		\if@verbose{
			\ps@typeout{psfig: including \@p@sfile \space }
		}\fi
		\special{epsf=\ps@predir\@p@sfile \space }
	}\fi
	\if@postlogfile
	    \special{\@postlogfileval \space } \fi
	\special{empty.ps /@isp {false} def}
}
\def\DvipsSpecials{
	\special{ps::[begin] 	\@p@swidth \space \@p@sheight \space
			\@p@sbbllx \space \@p@sbblly \space
			\@p@sbburx \space \@p@sbbury \space
			startTexFig \space }
	\if@clip{
		\if@verbose{
			\ps@typeout{(clip)}
		}\fi
		\special{ps:: doclip \space }
	}\fi
	\if@angle
		\if@verbose{
			\ps@typeout{(clip)}
		}\fi
		\special {ps:: \@p@sangle \space rotate \space} 
	\fi
	\if@prologfile
	    \special{ps: plotfile \@prologfileval \space } \fi
	\if@decmpr{
		\if@verbose{
			\ps@typeout{psfig: including \@p@sfile.Z \space }
		}\fi
		\special{ps: plotfile "`zcat \@p@sfile.Z" \space }
	}\else{
		\if@verbose{
			\ps@typeout{psfig: including \@p@sfile \space }
		}\fi
		\special{ps: plotfile \@p@sfile \space }
	}\fi
	\if@postlogfile
	    \special{ps: plotfile \@postlogfileval \space } \fi
	\special{ps::[end] endTexFig \space }
}
\def\psfig#1{\vbox {
	\ps@init@parms
	\parse@ps@parms{#1}
	\compute@sizes
	\ifnum\@p@scost<\@psdraft{
		\PsfigSpecials 
		\vbox to \@p@srheight sp{
			\hbox to \@p@srwidth sp{
				\hss
			}
		\vss
		}
	}\else{
		\if@draftbox{		
			\hbox{\fbox{\vbox to \@p@srheight sp{
			\vss
			\hbox to \@p@srwidth sp{ \hss 
			 \hss }
			\vss
			}}}
		}\else{
			\vbox to \@p@srheight sp{
			\vss
			\hbox to \@p@srwidth sp{\hss}
			\vss
			}
		}\fi

	}\fi
}}
\psfigRestoreAt
\setDriver
\let\@=\LaTeXAtSign

\begin{document}

\title {The extinction law in high redshift galaxies\footnote{
       Based on observations made with the NASA/ESA Hubble Space Telescope. 
       The Space Telescope Science Institute is operated by the Association of
       Universities for Research in Astronomy, Inc. under NASA contract 
       NAS 5-26555.}}


\author{J.A. Mu\~noz$^1$, E.E. Falco$^2$, C.S. Kochanek$^3$,
B.A. McLeod$^4$ and E. Mediavilla$^{5}$}

\bigskip
\affil{$^{1}$Departamento de Astronom\'{\i}a y Astrof\'{\i}sica, Universidad
       de Valencia, E-46100 Burjassot, Valencia, Spain}
\affil{$^{2}$F. L. Whipple Observatory, Smithsonian Institution, P. O. Box 97,
  Amado, AZ 85645, USA}
\affil{$^{3}$Department of Astronomy, The Ohio State University, Columbus,
   OH 43210, USA}
\affil{$^{4}$Harvard-Smithsonian Center for Astrophysics, 60 Garden St.,
  Cambridge, MA 02138, USA}
\affil{$^{5}$Instituto de Astrof\'{\i}sica de Canarias, E-38200 La Laguna,
       Tenerife, Spain}

\affil{email: jmunoz@uv.es}


\begin{abstract}
We estimate the dust extinction laws in two intermediate redshift 
galaxies.  The dust in the lens galaxy of LBQS~1009--0252, which 
has an estimated lens redshift of $z_l\simeq 0.88$, appears to be 
similar to that of the SMC with no significant feature at 2175\AA.
Only if the lens galaxy is at a redshift of $z_l\simeq 0.3$,
completely inconsistent with the galaxy colors, luminosity or
location on the fundamental plane, can the data be fit with a 
normal Galactic extinction curve.
The dust in the $z_l=0.68$ lens galaxy for B~0218+357, whose
reddened image lies behind a molecular cloud, requires a very flat
ultraviolet extinction curve with (formally) $R_V=12 \pm 2$.  
Both lens systems seem to have unusual extinction curves by
Galactic standards.
\end{abstract}


\section{Introduction}
Precise measurements of extinction curves are almost
exclusively limited to the Galaxy, the LMC and the SMC, because at greater
distances it becomes impossible to obtain the photometry or
spectroscopy of individual stars needed for accurate extinction
law measurements.  Galactic extinction curves are well fitted by parameterized
models with $R_V\simeq 3.1$ ($A_\lambda\equiv R_\lambda E(B-V)$), e.g.
Savage \& Mathis 1979, Fitzpatrick \& Massa 1988, Cardelli, Clayton \&
Mathis 1989, hereafter CCM), although lines of sight with dense molecular gas can show much
higher values (e.g. Jenniskens \& Greenberg 1993) for a total range
of $2.1 < R_V < 5.8$ (see e.g. Draine 2003 and references therein).  In
the Galaxy and most of the Large Magellanic Cloud (LMC), the changes in
the extinction law are strongly correlated with changes in the width and
amplitude of the 2175\AA\ feature in the extinction curve.  The extinction
curve of the Small Magellanic Cloud (SMC) and some regions of the LMC is
very different from the typical Galactic extinction law in having a
far weaker or nonexistent 2175\AA\ feature (e.g. Misselt et al. 1999, Gordon \etal\ 2003).
Physically, the extinction law depends on the mean size and composition
of the dust grains along the line of sight (e.g. Clayton \etal\ 2003, 
Draine \& Malhotra~1993, Rouleau et al. 1997), so it should not be surprising
 that it varies with the environment.  

There are only fragmentary data on the extinction curves in other local
galaxies.  A range of $R_V$ are found in M31, and the variations
may be correlated with the local metallicity
(Hodge \& Kennicutt 1982, Iye \& Richter 1985).  There is evidence
that the extinction curves of early-type galaxies are steeper functions
of $\lambda^{-1}$, but there are no cleanly measured extinction curves
(Warren-Smith \& Berry 1983, Brosch \& Loinger 1991, Goudfrooij et al. 1994).
At least in the optical (I-band through B-band), Riess et al. (1996)
used Type Ia supernovae to show that the extinction curves of nearby
galaxies were consistent with a mean optical extinction curve having
$R_V=2.6\pm0.3$.
In short, aside from one spiral galaxy and two irregulars (the Galaxy,
the LMC and the SMC) we have few quantitative measurements of dust
properties.

The galaxies, that have been studied so far, are not a representative sample of
galaxies or environments. Moreover, all the physics governing dust
properties (metallicity, star formation and evolution rates, radiation
backgrounds) evolve strongly with redshift, so we would expect the
properties of the dust to evolve with redshift.
Evolution
in the mean extinction law with redshift would be a crucial systematic
uncertainty in studies of Type Ia supernovae to constrain the
cosmological model (see, e.g., Perlmutter et al. 1999), since
extinction modifies the apparent stretch of the light curves (Nugent et al.~2002)
.
Extinction laws are also required for models of galaxy evolution (e.g., in
semi-analytic models, Silva et al.~2001, or population synthesis models,
Gordon et al. 1997), for estimates of star formation rates in individual
galaxies (e.g. Pettini et al. 1998; Meurer et al. 1999) or to construct
a global history of star formation (e.g. Madau et al. 1998; Steidel et
al. 1999).  Extinction also affects the light curves of $\gamma$-ray
bursts (e.g. Price et al. 2001, Jha et al.~2001), and deriving the
extinction law from afterglows requires theoretical assumptions about
the intrinsic spectrum of the burst.  With the increasing need for
extinction corrections at higher redshifts, it would be wise to
obtain more quantitative measurements of dust properties at similar
redshifts.   To make any progress,
we need a probe of extinction which has the precision of local stellar
measurements and works at $z=1$ just as well as at $z=0$.

Gravitational lenses provide a unique tool to study the extinction properties
of high redshift galaxies. In most of the $\sim80$ known lens galaxies
we see 2 or 4 images of a background AGN
produced by the deflection of light by a foreground lens galaxy. 
When each image's light traverses the lens galaxy, it is extinguished by
the dust at that position.  When the dust is not uniform, the amount of 
extinction is different for each image, whose observational
signature is that the flux ratios of the images depend on wavelength
(Nadeau \etal\ 1991).  Extinction curves have been estimated for
several systems using optical and infrared flux ratios 
(Nadeau \etal\ 1991, Jaunsen \& Hjorth 1997, Motta \etal\ 2002,
Wucknitz \etal\ 2003), and Falco \etal\ (1999) made a general survey of dust
properties using the available lens photometry.  
Unfortunately, ground-based observations can study the region
around the 2175\AA\ feature only for the highest redshift lenses. The
feature is redshifted to wavelengths above the atmospheric cutoff (3500 \AA)
for $z_l>0.6$, and is easily studied only for $z_l\simeq 1$.  Most
photometry from Hubble Space Telescope (HST) observations is limited
to the V, I and H-bands, since the observations were designed to study
the lens and host galaxies rather than extinction.  
In this paper we used near-UV observations with the Hubble Space Telescope (HST)
to study the extinction law of two gravitational lenses near the 2175\AA\ 
feature.  We summarize the observations in \S2 and the results in \S3. 

\section{Observations}

From the Falco \etal\ (1999) lens extinction survey, we selected 3
lenses with significant extinction whose redshifted 2175 \AA\ feature
would lie at longer wavelengths than the Lyman limit of the source
quasar (i.e. $2175(1+z_l)$ \AA $> 912 (1+z_s)$ \AA).
We used a total of 10 orbits to observe B~0218+357 (6 orbits), 
LBQS~1009--0252 (2 orbits) and Q~2337+0305 (2 orbits).

Table 1 shows a log of our WFPC2 observations.  Each image was composed of
dithered but not CR-split sub-exposures.  The original observing 
request was reduced, and our subsequent decision to obtain data for all the proposed
targets forced us to use only 2 sub-exposures for most of the observations.  
Unfortunately, for  Q~2237+0305, we found the short integrations insufficient 
to produce a useful extinction measurement and we include only the
photometric results for this system.
The sub-exposures were combined using standard methods (e.g. as in Leh\'ar
\etal\ 2000) but with 
additional manual masking to control the cosmic rays.  Table 2 presents
the magnitude measurements for all three systems. 

\section{Analysis and Discussion}

If $m_0(\lambda)$ is the intrinsic
QSO spectrum expressed as magnitudes at observed wavelength $\lambda$,
then the spectrum of lensed image $i$, $m_i(\lambda)$, is
\begin{equation}
m_i(\lambda)=m_0(\lambda) - M_i + E_i \,\,R\left(\frac{\lambda}{1+z_l}\right)
\end{equation}
 where $M_i$ and $E_i=E(B-V)$ are the
magnification and extinction of image $i$, and $ R(\lambda/(1+z_l))$
is the extinction curve redshifted to the lens redshift $z_l$.  By
measuring the magnitude differences as a function of wavelength for
each image pair (\eg, A and B)
\begin{equation}
m_B(\lambda)-m_A(\lambda)=\Delta M +\Delta E\,\,R\left(\frac{\lambda}{1+z_l}\right)
\end{equation}
we can measure the relative magnifications $\Delta M=M_B-M_A$, extinction
differences $\Delta E(B-V)=E_B(B-V)-E_A(B-V)$, and the mean extinction curve
R(V) without needing to know the intrinsic spectrum $m_0(\lambda)$.  We
assume that the shape of the source spectrum does not vary with time, 
that there is no wavelength dependence $M_i(\lambda)$ to the magnification 
due to microlensing, and that extinction curve is the same for all
images.  We will discuss these assumptions further below. 

We used a standard $\chi^2$ statistic to fit the model to the 
measurements and to determine $\Delta M$, $\Delta E$, $R_V$ and 
their uncertainties.  We used either the Cardelli \etal\ (1989) 
parameterized models for the Galactic extinction curve or the
Fitzpatrick \& Massa (1990) model with its parameters set to
the values found by Gordon \etal\ (2003) for the average extinction
in the SMC.  The Galactic models have a strong 2175\AA\ feature
while the SMC models do not.
We also attempted to determine the dust redshift $z_d$ by varying
the lens redshift $z_l$ in our fits (Jean \& Surdej 1998, Falco \etal\
1999).  If the wavelength dependence of the flux ratios is due to 
extinction, then we should find that $z_d$ is consistent with $z_l$.
Table 3 shows the results for these parameters. 

From the colors of the lens galaxy in LBQS~1009-0252 or its location on
the fundamental plane, we estimated that the lens redshift is 
$z_l=z_{FP}=0.88^{+0.04}_{-0.11}$ (Kochanek \etal\ 2000).  To date, these
estimated redshifts have always been confirmed by subsequent spectroscopic
measurements.  An example at similar redshift and with similar photometric
data is HE1104--1805, where the Lidman et al. (2000)
spectroscopic redshift of $z_l=0.73$ exactly matched the prior prediction
of $z_{FP}=0.73 \pm 0.04$ from the fundamental plane.  
Fig.~1 shows the magnitude differences as a function of
the observed wavelength, and there is no sign of a 2175\AA\ feature at the
$z_{FP}=0.88$ 
redshifted wavelength of $\lambda^{-1}\simeq 2.45\mu$m$^{-1}$.  For
$z_l=z_{FP}$ the best fit with a Galactic extinction curve
has $\chi^2=32.7$.  In these fits we included the H-band flux ratio
from Leh\'ar \etal~(2000), although the result changes little if it
is excluded.  Galactic dust is permitted if the redshift estimate
is wrong, as we find a perfect fit ($\chi^2=0.14$) with a Galactic
extinction law for $z_d=0.31\pm0.09$.  With this low redshift, the
2175\AA\ feature lies outside the range of our data (see Fig.~1).   
This is very unlikely given the colors and structural properties
of the lens galaxy because it is almost impossible for a lower
redshift galaxy to mimic the colors and structure of a higher
redshift early-type galaxy (see Leh\'ar \etal\ 2000 and Kochanek \etal\
2000).  

The data are, however, very well fitted ($\chi^2=0.65$) by an SMC extinction 
curve redshifted to the expected $z_l=z_{FP}=0.88$ (see Fig.~1).
The SMC extinction curves lack the features needed to estimate
a dust redshift, and we can find a good fit for almost any lens redshift
($0 \lesssim z_d \lesssim 2$ at 2$\sigma$).   Chromatic microlensing can
also produce wavelength-dependent flux ratios (see e.g. Yonehara \etal\ 1999),
but the lack of significant changes in the V and I flux ratios between 1999.01 and
1999.11 argues against a significant contribution from microlensing.  In order
to obtain such a rapid change in the flux ratio with wavelength using
microlensing, the source would have to lie in a highly magnified region 
of the microlensing magnification pattern where the time scales would be
relatively short.  This is easily checked by comparing the emission line
and continuum flux ratios in a spectrum, since extinction has the same
effect on both spectral components while microlensing primarily affects
the continuum fluxes (e.g. Wucknitz \etal\ 2003).

The second lens, B~0218+357, has a known spectroscopic redshift of $z_l=0.6847$
(Browne \etal\ 1993, Stickel \& Kuhr 1993).  We combined the 6 WFPC2 flux
ratios with the radio flux ratio of $m_B-m_A=1.40\pm0.03$ (Biggs \etal\ 1999).
The radio flux
ratio provides a direct constraint on the true magnification ratio $\Delta M$
because it is unaffected by extinction.  The magnitude differences (see Fig.~2)
show a feature at the redshifted wavelength
of the 2175\AA\ feature, but the overall pattern does not correspond to a 
standard $R_V=3.1$ Galactic extinction curve.  The data are well fitted by the
Cardelli et al.~(1989) model ($\chi^2=0.72$) for $R_V=12\pm2$.  This
is an extrapolation well past the upper values of $R_V \simeq 5$ actually
observed in the Galaxy (see e.g. Clayton \etal\ 2003 and references therein), 
but it maintains the overall structure of the high
$R_V$ extinction curves observed in the Galaxy.  In this case, the dust redshift
of $z_d=0.70\pm0.06$ is consistent with the spectroscopic redshift.
Moreover, for B~0218+357 we did not include the Leh\'ar \etal~(2000) photometry,
yet the model passes almost exactly through the H-band point lying between
the radio and optical data.  This is consistent with local observations that
the near-IR extinction curve is universal (e.g. Martin \& Whittet 1990).

B~0218+357 is known to lie behind a molecular
cloud from the existence of molecular absorption lines in the radio continuum
(see e.g. Combes \& Wiklind 1998). Given $A_V\sim4$~mag, an atomic Hydrogen 
column density of $N(H_I)\sim10^{21}$~cm$^{-2}$ (Carilli, Rupen \& Yanny 1993) and
molecular Hydrogen column density of between $N(H_2)\sim2\times10^{22}$~cm$^{-2}$ (
Wiklind \& Combes 1995; Gerin et al. 1997) and 
$5\times10^{23}$~cm$^{-2}$ (Wiklind \& Combes 1995; Combes \& Wiklind 1997;
see also Menten \& Reid 1996), we find a gas to dust ratio of 
$N_H/A_V\sim (1-25)\times 10^{22}$~cm$^{-2}$~mag$^{-1}$. 
The uncertainties are driven by the large range of the molecular column
density measurements.  
Gas-to-dust ratios in the Galaxy have an average value of
$N_H/A_V\sim 1.87\times 10^{21}$~cm$^{-2}$~mag$^{-1}$ (for $R_V=3.1$,
Bohlin \etal\ 1978) with a remarkable small scatter $\sim 30\%$. At higher
 $R_V$ the value of $N_H/A_V$ appears to be greater than the average, but the 
highest value measured, corresponding to $\rho$ Ophiuchus ($R_V=4.2$), is only
twice the average (see e.g. Kim \& Martin 1996).
Further study will be needed to understand a potential relation between the 
extreme value of $R_V\sim12$ with the huge measured $N_H/A_V$ ratio for
B~0218+357.

We explored whether adding dust with a different extinction law in front of the 
bluer image, B~0218+357B, would allow solutions with less extreme extinction curves.
These models are under constrained, so we computed models for fixed
ratios of the extinction between the two images ($E_B(B-V)=0.25 E_A(B-V)$, 
$0.50 E_A(B-V)$ and $0.75 E_A(B-V)$).  The results for a particular test,
using a standard Galactic $R_V=3.1$ extinction law for the B image, are
shown in Fig.~3.  As we increase the extinction for image B, we find
fits with lower values of $R_V$ for image A.  This behavior is 
generic over a wide range of choices ($2 < R_V^B < 5$) for the B image 
extinction curve. In the final analysis, however, this approach does not seem
to be a plausible solution because models with enough extinction in the
B image to allow significantly smaller values of $R_V^A$ also imply an
intrinsic source spectrum with an inverted 2175\AA\ feature.  This is
not physically plausible, so it is unlikely that the B image can suffer
significant extinction. These conclusions do not change if we assume an SMC-like
 extinction law for image B rather than a Galactic law. 

In summary, both the LBQS~1009--0252 and B~0218+357 lens systems seem to have 
unusual extinction curves.  This is not true of all lenses. Falco 
et al.~(1999) found several systems with extinction curves reasonably
consistent with standard Galactic models, and Motta \etal\ (2002) conclude
that the lens SBS~0909+532  is fitted well by a standard $R_V=2.1\pm0.9$
extinction law.
There is one caveat on our results, arising from the fact
that we ignored the possibility of wavelength dependent magnifications due
to microlensing from the stars in the lens galaxy (see e.g. Yonehara \etal\ 
1999).  Because the deviations
from normal extinction curves are large, both systems would have to have
large microlensing magnifications and correspondingly small source sizes.
At least for LBQS~1009--0252 this can be checked by studying the flux
ratios of the images in the emission lines and the continuum of the quasars,
where they should be the same if dust is responsible for the wavelength
dependence of the flux ratios and very different if microlensing is 
responsible.  Unfortunately, the source in B~0218+357 is a BL~Lac object
and lacks strong optical emission lines.  The flux ratios should also show
time variability on time scales of 1--10~years if the cause is microlensing,
and the lack of significant changes between our present results and 
Leh\'ar et al.~(2000) argues against microlensing.

Our results confirm the unique value of gravitational lenses for studying
extinction laws at cosmological distances.  Since the number of known 
lenses is increasing rapidly, the only barrier to further studies of
extinction laws is the need for near-UV HST observations to study the
region near the 2175\AA\ feature.

\bigskip
\noindent Acknowledgments:
JAM is a {\it Ram\'on y Cajal Fellow} from the MCyT of Spain.
EEF, CSK and BAM were supported in part by the Smithsonian Institution.
These observations were obtained as part of HST grant GO-8252.  CSK
is supported by NASA ATP grant NAG5-9265.


\newpage

\begin{figure}
\centerline{\psfig{figure=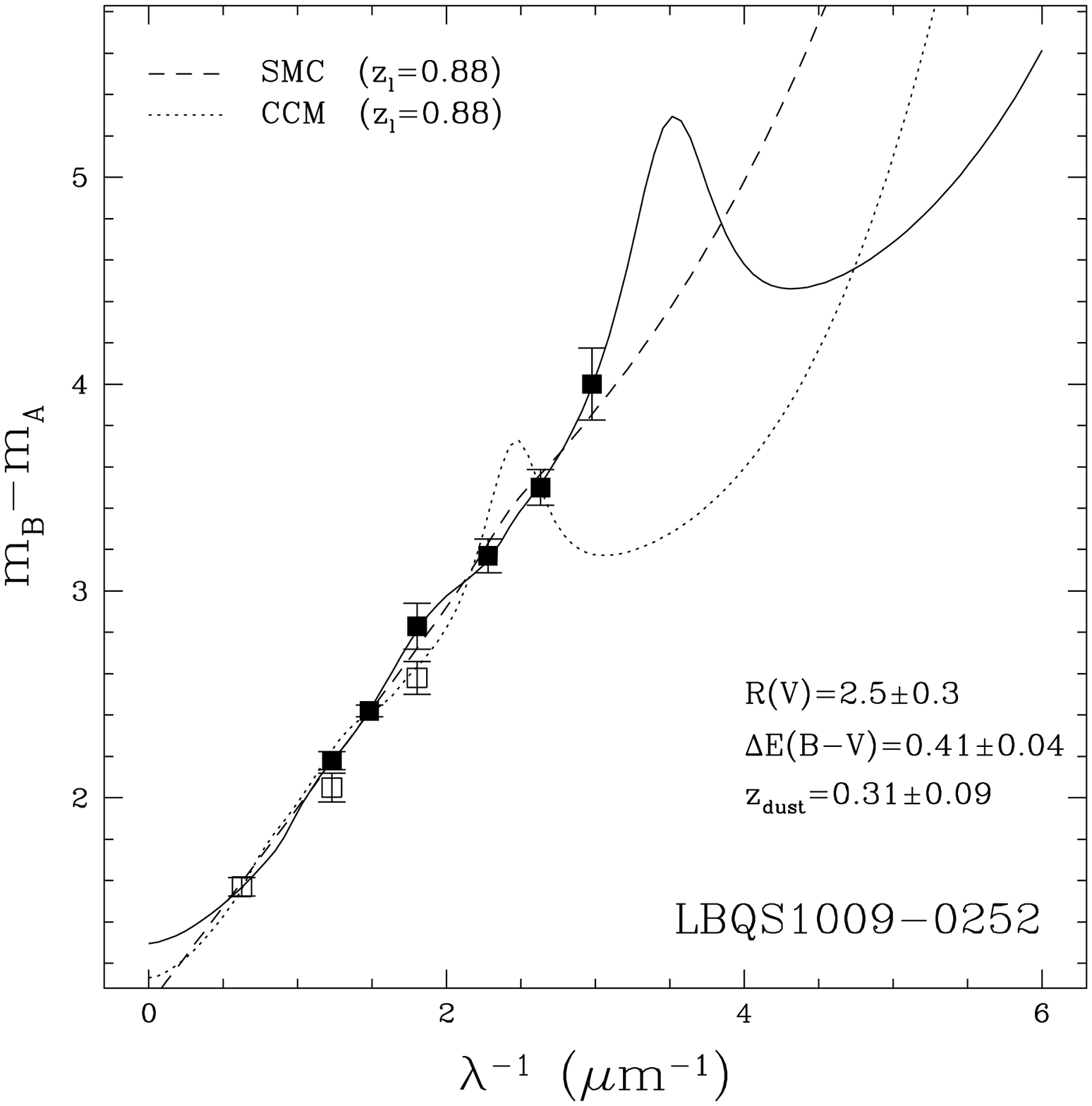,height=5.5in}}
\caption{The magnitude difference as a function of observed wavelength 
for LBQS~1009--0252.  The filled squares are the measurements from these
observations and the open squares are the earlier measurements from
Leh\'ar et al.~(2000). The solid line shows the best fit Galactic (CCM)
extinction curve, where a redshift of $z_d\simeq 0.3$ is required to
avoid predicting a 2175\AA\ feature in the data
(the dotted line shows a CCM extinction curve for a fixed $z_l=0.88$).
The dashed line
shows the best fit SMC extinction law at the best estimate for the
lens redshift ($z_l=z_{FP}=0.88$) from the color, luminosity and location
on the fundamental plane of the lens galaxy.  The infrared data from Leh\'ar
\etal~(2000) were included in the fits, although the result changes
little if we use only the data from the current observations. }
\end{figure}

\begin{figure}
\centerline{\psfig{figure=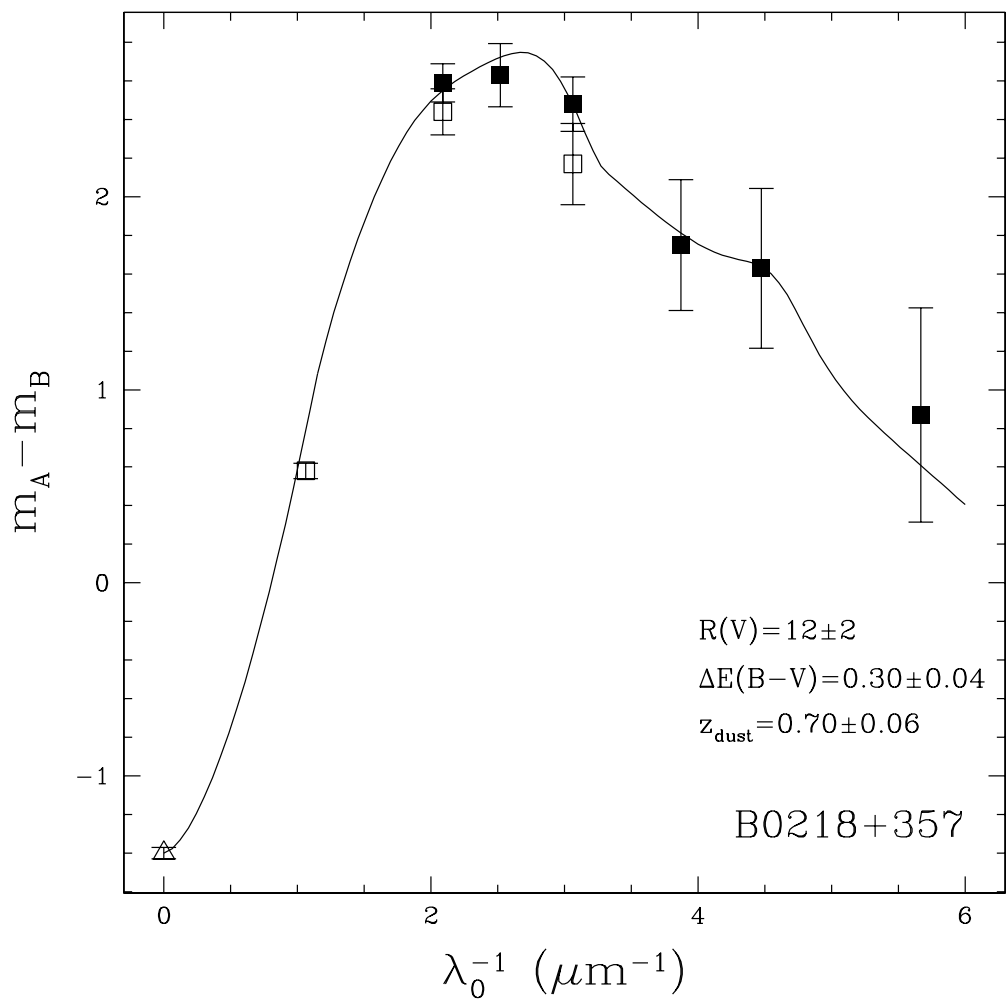,height=5.5in}}
\caption{ The magnitude difference as a function of the rest frame
wavelength for B~0218+357.  The filled squares are the measurements from these
observations, the open squares are the earlier measurements from
Leh\'ar et al.~(2000), and the triangle is the 
radio flux ratio (Biggs \etal\ 1999).  
The solid line shows the best fit Galactic (CCM) extinction curve. 
The Leh\'ar \etal~(1999) photometry (open squares) was not included in the
fit, yet the model passes almost exactly through the H-band 
($\lambda_0^{-1}=1.06 \mu m^{-1}$) point lying between the radio
and optical data.}
\end{figure}

\begin{figure}
\centerline{\psfig{figure=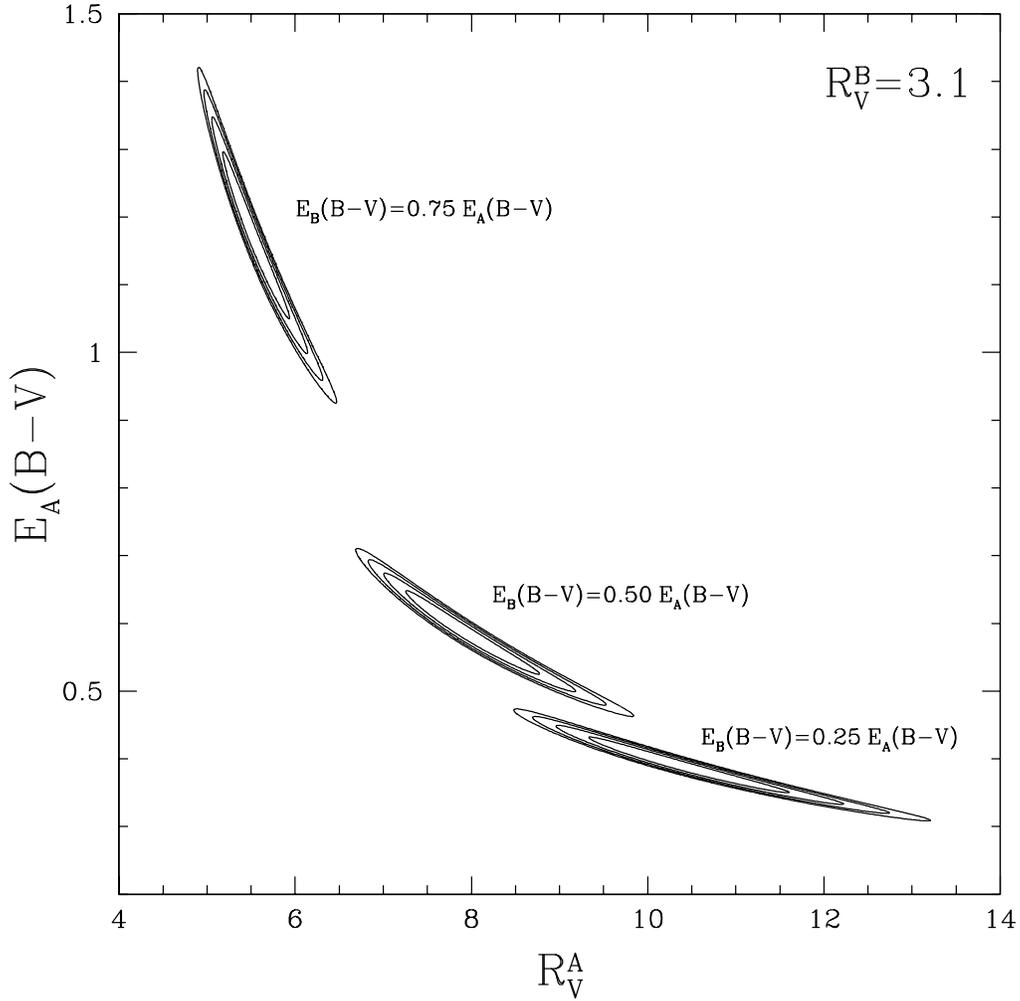,height=5.5in}}
\caption{
Examples of models for B~0218+357 that include extinction of the B image
with a different ($R_V=3.1$) extinction curve than is used for the A
image.  The $\chi^2$ contours are 
shown from $\chi^2_{min}+1$ to $\chi^2_{min}+4$ assuming
that $E_B(B-V)$ was fixed to be 0.75, 0.50 and 0.25 times 
$E_A(B-V)$.}
\end{figure}

\newpage

\begin{deluxetable}{lccrc}
\footnotesize
\tablewidth{0pt}
\tablecaption{Log of WFPC2 Observations}
\tablehead{
\colhead{TARGET} & \colhead{DATE-OBS} & \colhead{FILTER} & \colhead{EXP} &
\colhead{POS. ANGLE} \\
       &  \colhead{(yyyy-mm-dd)}& & \colhead{(sec)}& \colhead{ (deg E of N)}\\
}
\startdata
B~0218+357      &2000-03-08    &F300W&     3x700 &  100.908\\
B~0218+357      &2000-03-08    &F380W&    6x1300 &  100.908\\
B~0218+357      &2000-03-08    &F439W&     3x700 &  100.908\\
B~0218+357      &2000-03-08    &F555W&    2x2300 &  100.908\\
B~0218+357      &2000-03-08    &F675W&     2x180 &  100.908\\
B~0218+357      &2000-03-08    &F814W&     2x120 &  100.908\\
LBQS~1009--0252 &1999-11-05    &F336W&     2x500 &  $-$29.5607\\
LBQS~1009--0252 &1999-11-05    &F380W&     2x350 &  $-$29.5607\\
LBQS~1009--0252 &1999-11-05    &F439W&     2x180 &  $-$29.5607\\
LBQS~1009--0252 &1999-11-05    &F555W&      2x60 &  $-$29.5607\\
LBQS~1009--0252 &1999-11-05    &F675W&      2x60 &  $-$29.5607\\
LBQS~1009--0252 &1999-11-05    &F814W&      2x60 &  $-$29.5607\\
Q~2337+0305     &1999-10-20    &F218W&     2x180 &  123.703\\
Q~2337+0305     &1999-10-20    &F255W&     2x160 &  123.703\\
Q~2337+0305     &1999-10-20    &F300W&     2x120 &  123.703\\
Q~2337+0305     &1999-10-20    &F336W&      2x60 &  123.703\\
Q~2337+0305     &1999-10-20    &F439W&     2x120 &  123.703\\
Q~2337+0305     &1999-10-20    &F555W&      2x60 &  123.703\\
Q~2337+0305     &1999-10-20    &F675W&      2x60 &  123.703\\
Q~2337+0305     &1999-10-20    &F814W&      2x60 &  123.703\\
\enddata
\end{deluxetable}

\newpage

\begin{figure}
\centerline{\psfig{figure=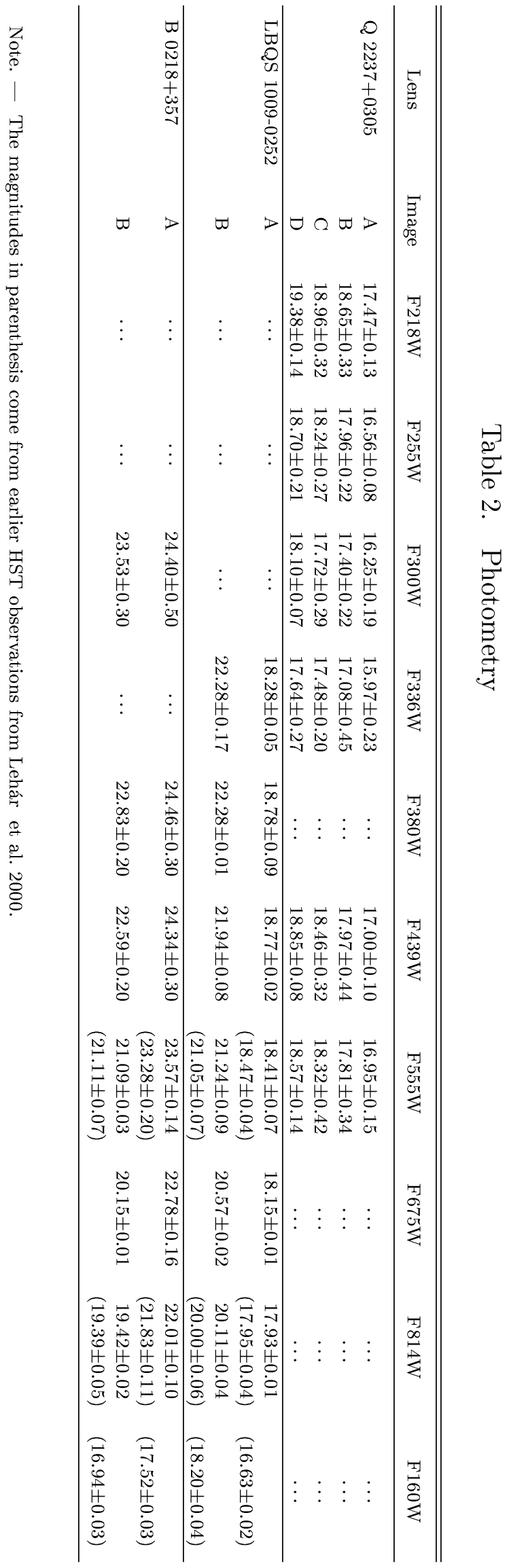,width=8in}}
\end{figure}

\newpage

\setcounter{table}{2}

\begin{deluxetable}{lccccccc}
\footnotesize
\tablewidth{0pt}
\tablecaption{Dust Properties using a CCM Extinction Curve}
\tablehead{ 
\colhead{Lens}  &\colhead{R(V)} & \colhead{$\Delta E(B-V)$} &
\colhead{z$_{\mbox{dust}}$} & \colhead{z$_l$} &\colhead{$\Delta M$} &\colhead{$\chi^2$}
 }
\startdata
LBQS1009-0252 & 2.5$\pm$0.3 & 0.41$\pm$0.04 & 0.31$\pm$0.09 & (0.88$^{+0.04}_{-0.11}$)$^*$ & 1.3$\pm$0.1 &0.14 \\
B0218+357     & 12$\pm$2    & 0.30$\pm$0.04 & 0.70$\pm$0.06 & 0.6847 & 1.4$\pm$0.3 & 0.72 \\
\tableline
\tableline
\enddata
\tablecomments{$^*$ The spectroscopic redshift for the lens LBQS~1009-0252 is 
still unknown. We use the Fundamental Plane redshift estimation from Kochanek 
\etal\ (2000).}
\label{extinc}
\end{deluxetable}


\begin{thebibliography}{}

\bibitem[Biggs et al.(1999)]{1999MNRAS.304..349B} Biggs, A.~D., Browne, 
I.~W.~A., Helbig, P., Koopmans, L.~V.~E., Wilkinson, P.~N., \& Perley, 
R.~A.\ 1999, \mnras, 304, 349 

\bibitem[Bohlin, Savage, \& Drake(1978)]{1978ApJ...224..132B} Bohlin, 
R.~C., Savage, B.~D., \& Drake, J.~F.\ 1978, \apj, 224, 132 

\bibitem[]{} Brosch, N. \& Loinger, F., 1991, A\&A, 249, 327

\bibitem[Browne, Patnaik, Walsh, \& Wilkinson(1993)]{1993MNRAS.263L..32B}
Browne, I.~W.~A., Patnaik, A.~R., Walsh, D., \& Wilkinson, P.~N.\ 1993,
\mnras, 263, L32


\bibitem[]{} Cardelli, J.A., Clayton, G.C. \& Mathis, J.S., 1989, ApJ, 345, 245

\bibitem[Carilli, Rupen, \& Yanny(1993)]{1993ApJ...412L..59C} Carilli, 
C.~L., Rupen, M.~P., \& Yanny, B.\ 1993, \apjl, 412, L59 

\bibitem[Clayton et al.(2003)]{2003ApJ...588..871C} Clayton, G.~C., Wolff, 
M.~J., Sofia, U.~J., Gordon, K.~D., \& Misselt, K.~A.\ 2003, \apj, 588, 871 

\bibitem[Combes \& Wiklind(1997)]{1997ApJ...486L..79C} Combes, F.~\&
Wiklind, T.\ 1997, \apjl, 486, L79

\bibitem[Combes \& Wiklind(1998)]{1998A&A...334L..81C} Combes, F.~\&
Wiklind, T.\ 1998, \aap, 334, L81

\bibitem[]{} Draine, B. \& Malhotra, S., 1993, ApJ, 414, 632

\bibitem[Draine(2003)]{2003ARA&A..41..241D} Draine, B.~T.\ 2003, \araa, 41, 
241 

\bibitem[Falco et al.(1999)]{1999ApJ...523..617F} Falco, E.~E.~et al.\ 
1999, \apj, 523, 617 

\bibitem[]{} Fitzpatrick, E.L,. \& Massa, D., 1988, ApJ, 307, 734

\bibitem[Fitzpatrick \& Massa(1990)]{1990ApJS...72..163F} Fitzpatrick, 
E.~L.~\& Massa, D.\ 1990, \apjs, 72, 163 

\bibitem[Gerin et al.(1997)]{1997ApJ...488L..31G} Gerin, M., Phillips,
T.~G., Benford, D.~J., Young, K.~H., Menten, K.~M., \& Frye, B.\ 1997,
\apjl, 488, L31

\bibitem[]{} Gordon, K.~D., Calzetti, D., \& Witt, A.~N.\ 1997, ApJ, 487, 625

\bibitem[Gordon et al.(2003)]{2003ApJ...594..279G} Gordon, K.~D., Clayton, 
G.~C., Misselt, K.~A., Landolt, A.~U., \& Wolff, M.~J.\ 2003, \apj, 594, 
279 

\bibitem[]{} Goudfrooij, P., de Jong, T., Hansen, L. \& Norgaard-Nielsen, H.U., 1994,
MNRAS, 271, 833

\bibitem[]{} Hodge, P.W. \& Kennicutt, R.C., 1982, 87, 264

\bibitem[]{} Iye, M. \& Richter, O.G., 1985, A\&A, 144, 471

\bibitem[Jaunsen \& Hjorth(1997)]{1997A&A...317L..39J} Jaunsen, A.~O.~\& 
Hjorth, J.\ 1997, \aap, 317, L39 

\bibitem[Jean \& Surdej(1998)]{1998A&A...339..729J} Jean, C.~\& Surdej, J.\ 
1998, \aap, 339, 729 

\bibitem[]{} Jenniskens, P. \& Greenberg, J.M., 1993, A\&A, 274, 439

\bibitem[]{} Jha, S., et al., 2001, ApJL, L155

\bibitem[Kim \& Martin(1996)]{1996ApJ...462..296K} Kim, S.~\& Martin, 
P.~G.\ 1996, \apj, 462, 296 

\bibitem[Kochanek et al.(2000)]{2000ApJ...543..131K} Kochanek, C.~S.~et 
al.\ 2000, \apj, 543, 131 

\bibitem[Leh{\' a}r et al.(2000)]{2000ApJ...536..584L} Leh{\' a}r, J.~et 
al.\ 2000, \apj, 536, 584 

\bibitem[Lidman et al. (2000)]{2000A&A...346L..62L} Lidman, C., Courbin, F.,
  Kneib, J.-P., Golse, G., Castander, F., \& Soucail, G., 2000, A\&A, 364, L62

\bibitem[]{} Madau, P., Pozzetti, L., \& Dickinson, M.\ 1998, ApJ, 498, 106

\bibitem[Martin \& Whittet(1990)]{1990ApJ...357..113M} Martin, P.~G.~\& 
Whittet, D.~C.~B.\ 1990, \apj, 357, 113 

\bibitem[Menten \& Reid(1996)]{1996ApJ...465L..99M} Menten, K.~M.~\& Reid,
M.~J.\ 1996, \apjl, 465, L99

\bibitem[]{} Meurer, G.~R., Heckman, T.~M., \& Calzetti, D.\ 1999, ApJ, 521, 64

\bibitem[]{} Misselt, K.A., Clayton, G.C., \& Gordon, K.D., 1999, ApJ, 515, 128

\bibitem[Motta et al.(2002)]{2002ApJ...574..719M} Motta, V., Mediavilla, E.,
Mu\~noz, J.A. ~et al.\ 2002, \apj, 574, 719 

\bibitem[Nadeau et al.(1991)]{1991ApJ...376..430N} Nadeau, D., Yee, 
H.~K.~C., Forrest, W.~J., Garnett, J.~D., Ninkov, Z., \& Pipher, J.~L.\ 
1991, \apj, 376, 430 

\bibitem[Nugent, Kim, \& Perlmutter(2002)]{2002PASP..114..803N} Nugent, P., 
Kim, A., \& Perlmutter, S.\ 2002, \pasp, 114, 803 

\bibitem[]{} Perlmutter, S., Turner, M. S. \& White, M. 1999, PhRvL, 83, 670

\bibitem[]{} Pettini, M. etal\ 1998, ApJ, 508, 539

\bibitem[]{} Price, P.~A.~et al.\ 2001, ApJL, 549, L7

\bibitem[]{} Riess, A.G., Press, W. \& Kirshner, R.P., 1996, ApJ, 473, 588

\bibitem[]{} Rouleau, F., Henning, T. \& Stognienko, R., 1997, A\&A, 322, 633

\bibitem[]{} Savage, B.D. \& Mathis, J.S., 1979, ARA\&A 17, 73

\bibitem[Silva et al.(2001)]{2001Ap&SS.276.1073S} Silva, L., Granato, 
G.~L., Bressan, A., Lacey, C., Baugh, C.~M., Cole, S., \& Frenk, C.~S.\ 
2001, \apss, 276, 1073 

\bibitem[Steidel et al.(1999)]{1999ApJ...519....1S} Steidel, C.~C., 
Adelberger, K.~L., Giavalisco, M., Dickinson, M., \& Pettini, M.\ 1999, 
\apj, 519, 1 

\bibitem[Stickel \& Kuhr(1993)]{1993A&AS..101..521S} Stickel, M.~\& Kuhr,
H.\ 1993, \aaps, 101, 521

\bibitem[]{} Warren-Smith, R.F. \& Berry, D.S., 1983, MNRAS, 205, 889

\bibitem[Wiklind \& Combes(1995)]{1995A&A...299..382W} Wiklind, T.~\&
Combes, F.\ 1995, \aap, 299, 382

\bibitem[Wucknitz, Wisotzki, Lopez, \& Gregg(2003)]{2003A&A...405..445W}
Wucknitz, O., Wisotzki, L., Lopez, S., \& Gregg, M.~D.\ 2003, \aap, 405,
445

\bibitem[Yonehara et al.(1999)]{1999A&A...343...41Y} Yonehara, A., 
Mineshige, S., Fukue, J., Umemura, M., \& Turner, E.~L.\ 1999, \aap, 343, 
41 

\end{thebibliography}
\end{document}